\documentclass[floatfix,12pt]{iopart}
\pdfoutput=1
\usepackage{iopams}
\usepackage{graphicx}
\usepackage{psfrag}
\usepackage{color}

\def\Id{{\mathbb I}}

\begin{document}
\title{Defect production in quench from current-carrying non-equilibrium state}
\author{Dragi Karevski}
\address{Institut Jean Lamour, dpt. P2M, Groupe de Physique Statistique, Universit\'e de Lorraine, CNRS, B.P. 70239, F-54506 Vandoeuvre les Nancy Cedex, France.}
\ead{\mailto{dragi.karevski@univ-lorraine.fr}}

\author{Rosemary J. Harris} 
\address{School of Mathematical Sciences, Queen Mary University of London, Mile End Road, London, E1 4NS, United Kingdom.}
\ead{\mailto{rosemary.harris@qmul.ac.uk}}

\begin{abstract}
We consider the defect production of a quantum system, initially prepared in a current-carrying non-equilibrium state, during its unitary driving through a quantum critical point.
At low values of the initial current, the quantum Kibble-Zurek scaling for the production of defects is recovered. However, at large values of the initial current, i.e., very far from an initial equilibrium situation, a universal scaling of the defect production is obtained which shows an algebraic dependence with respect to the initial current value. These scaling predictions are demonstrated by the exactly solvable Ising quantum chain where the current-carrying state is selected through the imposition of a Dzyaloshinskii-Moriya interaction term. 
\end{abstract}

\section{Introduction}
When a system is driven adiabatically close to a critical point or through a gapless phase, the divergence of its intrinsic relaxation time leads to a complete breakdown of the adiabatic condition, no matter how slow the driving is. In the vicinity of the gapless point the dynamics switches from an adiabatic to a sudden quench regime.   Consequently, as we drive the system from its initial ground state closer and closer to the gapless point, the departure from the adiabatic evolution induces transitions towards excited states which are seen as a proliferation of topological defects. This proliferation ultimately generates a final state which differs significantly from the naively expected ground state associated to the final Hamiltonian. 
For a slow driving rate, generalizing to the quantum situation the classical Kibble-Zurek Mechanism (KZM) \cite{KZMclassical}, the density of such defects is expected to be given by a power-law function of the driving rate with an exponent related to the quantum critical point exponents \cite{ZuDoZo05,Da05,Po05,Dz05,BaPo08}. 

The scaling argument goes as follows. For a linear ramping through an isolated quantum critical point controlled by the parameter $\epsilon(t)\sim t/\tau_Q$, the system, after following adiabatically the ramping far away from the critical point, will suddenly freeze out when it gets close enough to the critical locus. This happens at a typical time scale $\tau_{KZ}$ which is deduced self-consistently by equating the intrinsic  relaxation time with the inverse of the instantaneous energy gap $\Delta(t)$ at $t=\tau_{KZ}$, where the typical scaling of the inverse gap, $\Delta^{-1}(\tau_{KZ})\sim |\epsilon(\tau_{KZ})|^{-z\nu}$, is set by the deviation from the critical point $\epsilon(\tau_{KZ})$  at that time. One obtains $\tau_{KZ}\sim \tau_Q^{z\nu/(1+z\nu)}$ and accordingly the associated typical length scale 
$\xi_{KZ}\sim \tau_{KZ}^{1/z}\sim  \tau_Q^{\nu/(1+z\nu)}$ which gives an estimate of the defect density as $n_{exc}\sim \xi_{KZ}^{-d}\sim \tau_Q^{-d\nu/(1+z\nu)}$. 
This scaling prediction, based on an adiabatic-sudden-adiabatic evolution scenario \cite{Da05}, has been tested numerically and by analytical means in  a great variety of models; see \cite{Dziarmaga2010,Polkovnikov2011} for extensive reviews.   
In particular, in a number of integrable models, the KZM can be derived exactly from the mapping to a set of independent two-level systems, each of them undergoing Landau-Zener (LZ)  anti-crossings \cite{LZ}. Integrating the LZ transition probabilities over all modes leads finally to the KZM prediction for the density of excitations \cite{Dz05,Dziarmaga2010,LZKZM}. Initially introduced for homogeneous systems, the KZM has since been generalized to inhomogeneous situations such as those generated by the release of a power-law confining potential \cite{CoKa10} or by the propagation of a domain wall or critical front \cite{CritFront}.
The KZM has also been used in spatially inhomogeneous situations to describe symmetry-breaking phase transitions in space \cite{KZspace}. 

In all these cases the non-equilibrium situation generated by the temporal variation of a Hamiltonian parameter is obtained from an initial equilibrium state (generally the ground state since the system is supposed to be at zero temperature).  However, many situations of physical and technological interest have to do with starting states that are intrinsically out-of-equilibrium, for example, situations where there is a macroscopic current flowing through the system as a result of a coupling with two different baths or reservoirs.  Starting the driving process from such an excited state will strongly affect the way the defects are generated especially when the dynamical system comes close to a gapless point.  If the initial current density is small enough, one expects to recover a density of defects which is governed by (almost) the usual KZM prediction.  However, it is possible that this prediction could break down completely at large initial current values. It is the aim of this paper to clarify the issue.

In a one-dimensional system, a current-carrying state can be prepared by the contact of the system at its boundaries with  reservoirs (or heat baths) at different chemical potentials (or temperatures). In such a case, the system steady state will in general be a statistical mixture, well described  close to equilibrium (small gradient of chemical potential or temperature) by a MacLennan-Zubarev density matrix  \cite{MacLennan}. Far away from equilibrium, there is no general prediction for the density matrix of such a steady state although some exact results in terms of Matrix Product States have recently been obtained for integrable models \cite{KarevskiNEQ,XXZNEQ}. 
Another generic situation where such a current-carrying state emerges asymptotically in time is the case of the relaxation of a one-dimensional system in which initially the left and right halves  have been set to different temperatures, or chemical potentials, and then glued together by a local coupling.  It was shown on quasi-free systems (such as the quantum $XY$ spin chain) that the steady state reached from this type of initial state by a unitary evolution is effectively described by a generalized Gibbs distribution of the form 
$e^{-\bar{\beta} (H - \lambda Y)}$ where $H$ is the Hamiltonian of the chain, $\bar{\beta}=(\beta_L+\beta_R)/2$ the average inverse temperature, $\lambda=(\beta_L-\beta_R)/\bar{\beta}$ the non-equilibrium driving force, and $Y$ a long-ranged conjugated current operator (which describes an infinite set of conserved quantities) \cite{Ogata}. Moreover, in one-dimensional critical systems it has been proven by conformal field theory techniques that the steady-state currents are universal, depending only on the central charges of the theories and on the external temperature or chemical potential bias \cite{DoyonBernard}. 
In the zero temperature limit  the generalized Gibbs  or MacLennan-Zubarev density matrix  
$\rho\sim e^{-\beta(H-\lambda Y)}$ reduces to the ground state projector $|GS_Y\rangle\langle GS_Y|$ associated to the effective Hamiltonian $H-\lambda Y$. This is basically the physical justification for the use of the Lagrange multiplier method developed in \cite{Antal97,Antal}. 
Specifically, to the Hamiltonian $H$ of the considered system one adds a term $-\lambda \hat{J}$ proportional to a given current operator $\hat{J}$ (associated to a given conserved quantity). The ground state of the effective Hamiltonian $H-\lambda \hat{J}$ will then 
carry a non-vanishing mean current $\langle \hat{J}\rangle\neq 0$ at sufficiently large values of $\lambda$.
This effective ground  state is then interpreted  as a non-equilibrium, current-carrying, state of the original model described by $H$. Such an effective approach is believed to capture, at least locally, the essential features of a system coupled to two different quantum reservoirs at sufficiently low temperature.  
{{In fact, it is known to yield an \emph{exact} description of steady states for energy transport in critical systems, where $\hat{J}$ becomes the total momentum operator \cite{DoyonBernard,Bhaseen15}.  This latter observation suggests that analysis of $H-\lambda \hat{J}$ is particularly relevant for studying scaling exponents.}

In the following we will apply this strategy to the Ising quantum chain to select a state carrying an energy density current. Given that state, we will drive the Ising chain through its quantum critical point and focus on the asymptotic defect generation.   Finally, after the presentation of the model and its explicit solution, we will give an LZ calculation leading to the density of defects and extract from it a general argument for the generation of defects in such current-carrying situations.

\section{Model}

We start by considering the transverse Ising Hamiltonian on a one-dimensional lattice of $L$ sites with periodic boundary conditions:
\begin{equation}
H=-\sum_l \sigma_l^x \sigma_{l+1}^x - \frac{h}{2} \sum_l  \sigma_l^z.
\end{equation}
Here the $\sigma_l$'s are the usual Pauli spin matrices and $h$ is a field favouring alignment in the $z$ direction.  For definiteness we take $h>0$.  It is straightforward to show that an energy current in this model can be defined as
\begin{equation}
\hat{J}=\sum_l \hat{J}_l = \frac{h}{4} \sum_l \left( \sigma_l^x \sigma_{l+1}^y - \sigma_l^y \sigma_{l+1}^x  \right) 
\end{equation}
with current conservation reflected by $[H,\hat{J}]=0$.  Following the approach outlined in the introduction, we argue that a current-carrying excited state of $H$ can be generated as the ground state of the effective Hamiltonian
\begin{equation}
H_J= H - \lambda \hat{J} \label{e:master}
\end{equation}
where the $\lambda \hat{J}$ term introduces an interaction of Dzyaloshinskii-Moriya form and, without loss of generality, we assume $\lambda \geq 0$. Such an effective Hamiltonian was previously studied in, e.g.,~\cite{Antal97, Eisler03,Das11b} and, as we recap below, can be brought into diagonal free-fermion form by a series of exact transformations.\footnote{Subtleties relating to the boundary conditions under these transformations are not expected to be relevant in the thermodynamic limit; see, e.g.,~\cite{Lieb61} for more detailed analysis.}

Firstly, using the standard Jordan-Wigner transformation~\cite{Jordan28}
\begin{eqnarray}
\sigma_l^+ &= c_l^\dagger \exp\left(i\pi \sum_{j<l} c_j^\dagger c_j\right) \\
\sigma_l^- &= \exp\left(- i\pi \sum_{j<l} c_j^\dagger c_j\right) c_l 
\end{eqnarray}
the complete Hamiltonian can be written as
\begin{eqnarray}
H_J =&-\sum_l \left[ \left(c_l^\dagger c_{l+1}^\dagger + c_l^\dagger c_{l+1} - c_l c_{l+1}^\dagger -c_l c_{l+1}\right) \right.  \nonumber \\
&+ \left. \frac{h}{2} \left(2 c_l^\dagger c_l -1\right) + \frac{\lambda hi}{2} \left( c_l^\dagger c_{l+1} + c_l c_{l+1}^\dagger \right)\right]
\end{eqnarray}
where the $c_l$'s are fermion operators.  The next step is a Fourier transform to wave fermions $\alpha_k$,
\begin{eqnarray}
c_l^\dagger &= \frac{1}{\sqrt{L}} \sum_k \alpha_k e^{ikl} \\ 
c_l &=  \frac{1}{\sqrt{L}} \sum_k \alpha_k^\dagger e^{-ikl},
\end{eqnarray}
which, after some manipulation using fermion anti-commutation rules, yields
\begin{equation}
\fl H_J=\sum_k \left\{ \left[h \lambda \sin k + (h + 2\cos k) \right] \alpha_k^\dagger \alpha_k  + i \sin k (\alpha_k^\dagger \alpha_{-k}^\dagger + \alpha_k \alpha_{-k} )+ \frac{h}{2} \right\}. \label{e:Fourier}
\end{equation}

Finally, we perform a Bogoliubov-type similarity transform~\cite{Bogoliubov58}
\begin{eqnarray}
\alpha_k &= \eta_k \cos \omega_k  - i \eta_{-k}^\dagger \sin \omega_k  \\
\alpha_{-k}^\dagger &= -i \eta_{k} \sin \omega_k  + \eta_{-k}^\dagger \cos \omega_k.  
\end{eqnarray}
Here $\omega_k$ is assumed odd in $k$, i.e., $\omega_{-k}=-\omega_k$ which implies that terms of the form $\eta_{-k} \eta_{k} h \lambda \sin k \times i \cos \omega_k \sin \omega_k$ (and conjugate) are also odd and cancel in the sum.  The off-diagonal terms in the remaining part are cancelled by the choice
\begin{equation} 
\tan2\omega_k= \frac{2\sin k}{h+2 \cos k}. \label{e:bog}
\end{equation}
Note that this is exactly the same Bogoliubov angle as in the $\lambda=0$ case which is, in fact, to be expected since the current operator commutes with the Hamiltonian.  With the natural choice that $2\omega_k$ is in the same quadrant as the point $(h+2 \cos k, 2\sin k)$ so that $\omega_k$ takes the sign of $k$, the Hamiltonian $H_J$ can then be written in diagonal form as
\begin{equation}
H_J= \sum_k \eta_k^\dagger \eta_k \left( h \lambda \sin k + \sqrt{(h+2\cos k)^2 + 4 \sin^2 k} \right) + \mathrm{const.}
\end{equation}
We observe immediately that $\lambda\neq0$ breaks the $k \leftrightarrow - k$ symmetry of the spectrum; see, e.g.,~\cite{Antal97}. 
Following~\cite{Eisler03}, we show in figure~\ref{f:phase} 
\begin{figure}
\centering
\includegraphics[width=0.8\textwidth]{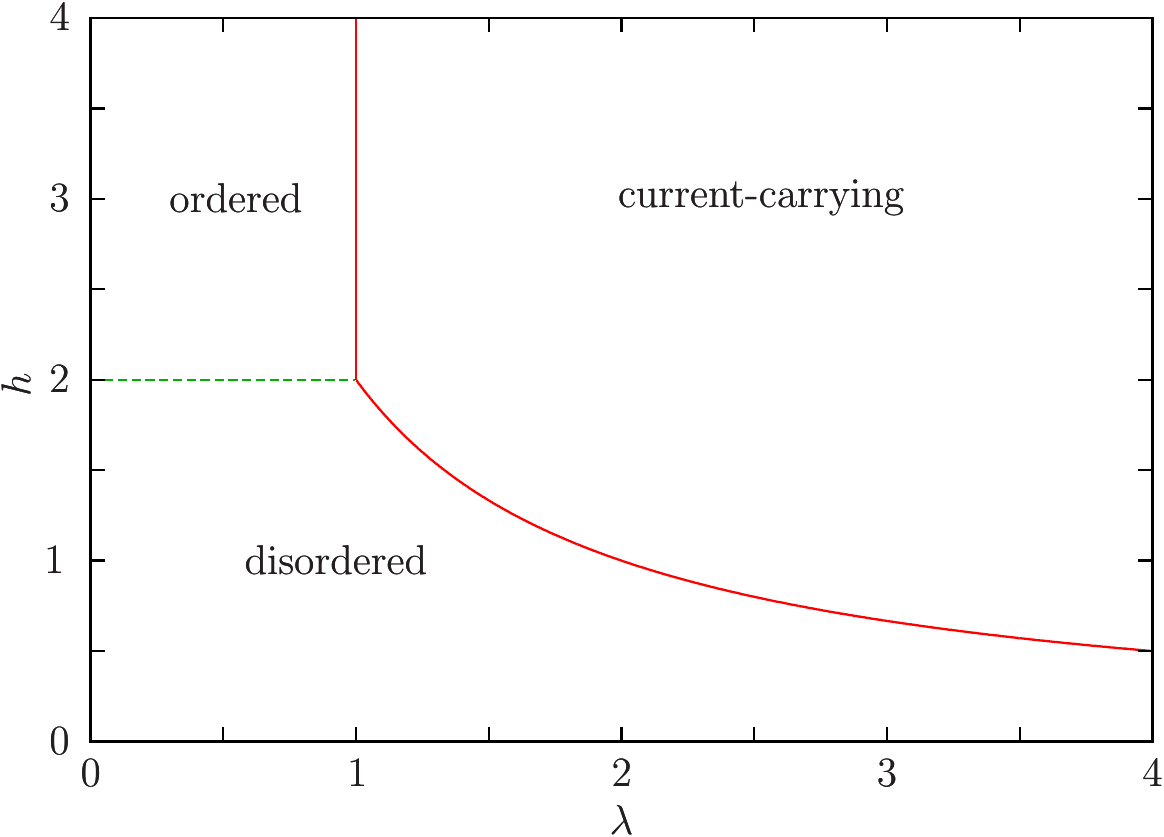}
\caption{Phase diagram corresponding to Hamiltonian $H_J$~\eref{e:master}.  The mean current is non-zero for $\lambda>\lambda_c(h)=\max(1,2/h)$.
In later sections we use this construction to consider a quench starting from a current-carrying initial state with $h>2$.}
\label{f:phase}
\end{figure}
the resulting phase-diagram in $h-\lambda$ space noting the slightly different parameterization of our Hamiltonian to that in the literature.
As indicated by the spectrum in figure~\ref{f:spec}, 
\begin{figure}
\centering
\includegraphics[width=0.8\textwidth]{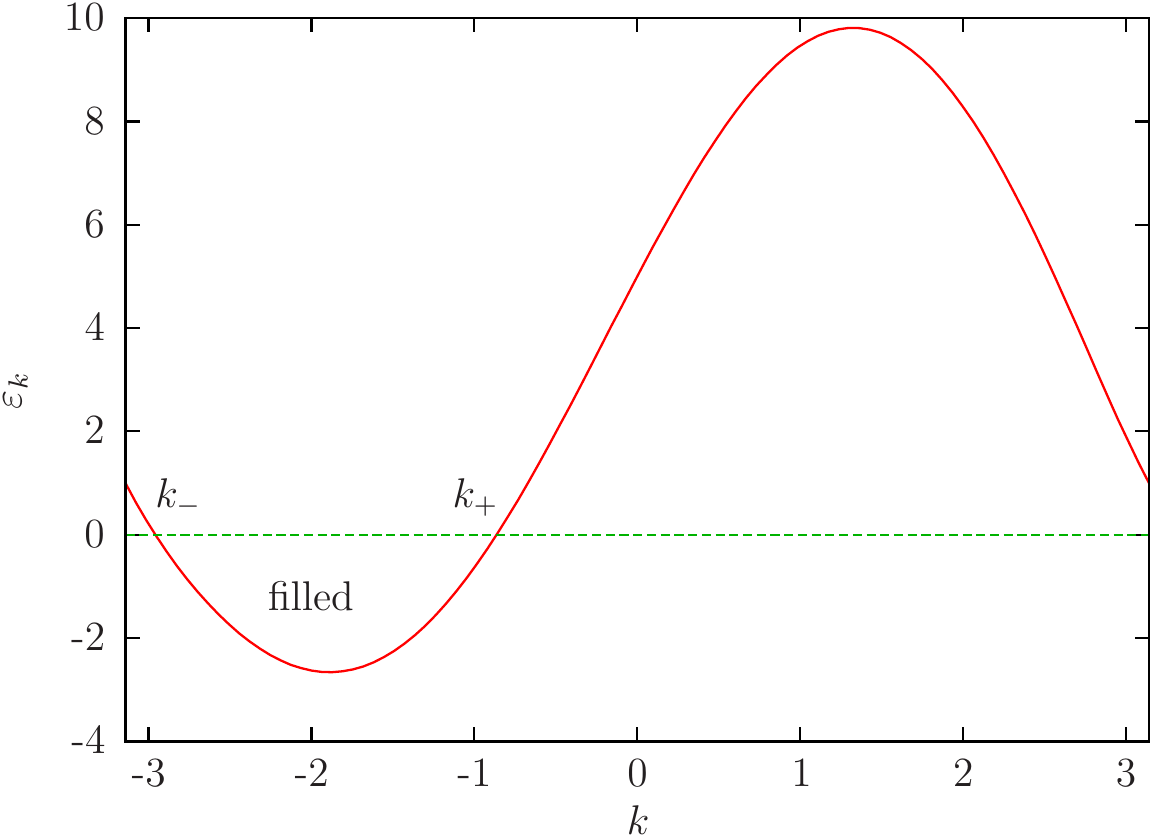}
\caption{Spectrum $\varepsilon_k=h \lambda \sin k + \sqrt{(h+2\cos k)^2 + 4 \sin^2 k}$ for $h=3$, $\lambda=2$ (solid red line) with zero energy line (dashed green) shown for comparison.  States between $k_-$ and $k_+$ are filled current-carrying modes.}
\label{f:spec}
\end{figure}
for $\lambda>\lambda_c(h)=\max(1,2/h)$ the ground state of $H_J$ consists of a band of filled current-carrying modes between $k_-$ and $k_+$ where
\begin{equation}
\cos k_\pm = \frac{-2 \pm \sqrt{(\lambda^2 h^2 -4)(\lambda^2-1)}}{h \lambda^2}. \label{e:qpm}
\end{equation}

In concluding this {section}, we emphasize that the dynamics is unaltered by the addition of the Dzyaloshinskii-Moriya term {(since $\lambda \hat{J}$ commutes with $H$)} but the effective Hamiltonian $H_J$ \textcolor{black}{(with its $\lambda$-dependent ground state)} can be used as a tool to generate a current-carrying initial state. Our programme in the following section is to analyse the creation of defects when the system starts from such a current-carrying initial state 
and is then quenched {across the critical line which, with our parameterization, is at $h=2$.  A related discussion on the properties of the entanglement entropy under a similar quench can be found in~\cite{Das11b}.  In fact, there is another critical line at $h=-2$ but as we restrict ourselves throughout to $h>0$, our quench protocol never crosses this.

\section{Calculation of defect production}

\subsection{Details of dynamics in Heisenberg picture}

In order to analyse the defect production it is helpful to follow the seminal paper of Barouch, McCoy and Dresden~\cite{BMD} and consider the dynamics from the Heisenberg viewpoint.   We start by writing~\eref{e:Fourier} as a sum over positive modes $p$,
\begin{equation}
H=\sum_{p=1}^{L/2} \tilde{H}_p,
\end{equation}
where
\begin{eqnarray}
\tilde{H}_p =&  (h+2\cos k)[ \alpha_k^\dagger \alpha_k + \alpha_{-k}^\dagger \alpha_{-k} ]
+ h\lambda\sin k [ \alpha_k^\dagger \alpha_k - \alpha_{-k}^\dagger \alpha_{-k} ] \nonumber \\
&+ 2i\sin k [\alpha_k^\dagger \alpha_{-k}^\dagger + \alpha_{k} \alpha_{-k} ] + h
\end{eqnarray}
with $k=({2\pi}/{L})p$. With the obvious choice of basis $\{|0\rangle,\alpha_k^\dagger \alpha_{-k}^\dagger |0\rangle, \alpha_k^\dagger |0\rangle, \alpha_{-k}^\dagger |0\rangle\}$ in which $|0\rangle$ is the vacuum state of the $\alpha$ fermions, we then have the $4\times4$ matrix representation
\begin{equation}
\fl
\tilde{H}_p =
\left(
\begin{array}{cccc}
h & 2i \sin k & 0 & 0 \\
- 2i \sin k & 4\cos k +3h & 0 & 0 \\
0 & 0 & h\lambda \sin k+2\cos k + 2h & 0 \\
0 & 0 & 0 &  -h\lambda \sin k+2\cos k + 2h \end{array}
\right). \label{e:Hmatrix}
\end{equation}
We note that the current-carrying states $\alpha_k^\dagger |0\rangle$ and $\alpha_{-k}^\dagger |0\rangle$ are completely decoupled from the other states and the diagonal structure of their submatrix indicates the conservation of current for constant field. 

The time evolution matrix, $U_p(t)$, in the Heisenberg picture obeys the ($\hbar=1$) equation 
\begin{equation}
i \frac{d}{dt} U_p(t) = U_p(t) \tilde{H}_p(t) \label{e:Hberg}
\end{equation}
with initial condition $U_p(t_0)=\Id$.  Here $\Id$ is the $4\times 4$ identity matrix and we have explicitly now included the time-dependence in $\tilde{H}_p(t)$ to allow for the time-dependent $h(t)$ which will be of interest in the following.  After the system has reached a current-carrying steady state corresponding to a non-zero $\lambda$, we consider a quench in $h(t)$ with the usual $\lambda=0$ dynamics so that the non-trivial part of~\eref{e:Hberg} reduces to an equation in the $2\times2$ basis $\{|0\rangle,|2\rangle=\alpha_k^\dagger \alpha_{-k}^\dagger |0\rangle\}$:
\begin{equation}
\fl
i \frac{d}{dt} 
\left(
\begin{array}{cc}
U_{11}(t) & U_{12}(t) \\
U_{21}(t) & U_{22}(t)
\end{array}
\right)
= 
\left(
\begin{array}{cc}
U_{11}(t) & U_{12}(t) \\
U_{21}(t) & U_{22}(t)
\end{array}
\right)
\times
\left(
\begin{array}{cc}
h(t) & 2i\sin k \\
-2i \sin k & 4\cos k + 3h(t)
\end{array}
\right)
\end{equation}
where we have suppressed the $p$ subscript in the matrix elements for notational brevity.  This system of coupled first-order differential equations, easily yields  decoupled second-order ones.  For example, we have
\begin{eqnarray}
\fl
i U_{11}'' &= h' U_{11} + h U_{11}' - (2i\sin k) U_{12}' \\
\fl &= h' U_{11} + h U_{11}'  + i(2i\sin k)[(2i\sin k) U_{11} + (4\cos k+3h) U_{12}] \\ 
\fl &= h' U_{11} + h U_{11}'  - (4 i\sin^2 k) U_{11} - 2\sin k (4\cos k+3h)\left[ \frac{iU_{11}'-hU_{11}}{-2i\sin k} \right] \\
\fl &= (4\cos k + 4h) U_{11}'+[h'-4i\sin^2 k + i(4 \cos k+3h) h]U_{11}
\end{eqnarray}
with initial condition $U_{11}(t_0)=1$ and $U'_{11}(t_0)=
-ih(t_0)$.
Similarly, one finds
\begin{equation}
iU_{12}'' = (4\cos k + 4h) U_{12}'+[3h'-4i\sin^2 k + i(4 \cos k+3h)h]U_{12}
\end{equation}
with initial condition $U_{12}(t_0)=0$, $U'_{12}(t_0)=2\sin k$.   The differential equations for $U_{21}$ and $U_{22}$ are identical to those for $U_{11}$ and $U_{12}$ respectively but with different initial conditions: $U_{21}(t_0)=0$, $U'_{21}(t_0)=-2\sin k$, $U_{22}(t_0)=1$, $U'_{22}(t_0)=-i[4 \cos k + 3 h(t_0)]$.

For certain choices of $h(t)$ the corresponding differential equations can be solved analytically (at least with the aid of a suitable computer algebra package) in terms of extremely tortuous combinations of hypergeometric/special functions.   However, our focus here is rather on using properties of the solutions to extract the scaling of the defect production when the system is quenched across the critical line.  Specifically, we first calculate the defect density starting from an initial state with $\lambda$ below the critical value (i.e., a steady state with zero current) and then demonstrate how the apparently simple change in the analysis required for a current-carrying initial state can lead to a dramatic change in the results.

We consider quenching the system from above the critical line ($h=2$) at some initial time $t_0$ to below the critical line at some final time $t_f$, according to a given smooth protocol \textcolor{black}{(with the shorthand definitions $h_0:=h(t_0)$ and $h_f:=h(t_f)$ now introduced)}.  \textcolor{black}{ For $\lambda<1$, since there is no current,  the system starts in the ground state of the $\eta$ fermions which we denote as
\begin{equation}
| \psi(t_0) \rangle = \prod_{k} |\tilde{0}_k(t_0)\rangle,
\end{equation}
where  the $|\tilde{0}_k(t_0)\rangle$ are the vacuum states 
associated to the diagonal fermions at time $t_0$ such that $\eta_k(\textcolor{black}{h_0})|\tilde{0}_k(t_0)\rangle=0$. }

To calculate the production of defects we need to consider the expectation of $\eta_k^\dagger(\textcolor{black}{h_f})\eta_k(\textcolor{black}{h_f})$  with respect to the time-evolved ground-state $|\psi(t_f)\rangle = U^\dagger(t_f) |\psi(t_0)\rangle$.   \textcolor{black}{To understand this, recall that at any time $t$ the system has a field value $h(t)$ and is diagonalized in terms of the operators $\eta_k^\dagger (h(t))\eta_k(h(t))$.  The associated (adiabatically expected) ground state is the vacuum state with respect to these fermions (since the single particle spectrum is positive). As a consequence, the number of defects is just the number of fermions on top of the instantaneous vacuum. At very low fields, $h\simeq 0$, the total number of defects $\sum_k \eta_k^\dagger(h)\eta_k(h)$ reduces to the kink number operator $\frac{1}{2}\sum_l (1-\sigma^x_l\sigma^x_{l+1})$,  see \cite{Dziarmaga2010,CoKa10} for more explanation.}  Since the dynamics is most easily expressed in terms of the $\alpha$ fermions we use the time-dependent version of the inverse Bogoliubov transformation to write
\begin{equation}
\fl
\eta_k^\dagger(\textcolor{black}{h_f})\eta_k(\textcolor{black}{h_f})= -is(\textcolor{black}{h_f})c(\textcolor{black}{h_f}) \alpha_{-k} \alpha_{k} + s(\textcolor{black}{h_f})^2 \alpha_{-k}\alpha_{-k}^\dagger +  c(\textcolor{black}{h_f})^2 \alpha_{k}^\dagger \alpha_{k} + is(\textcolor{black}{h_f})c(\textcolor{black}{h_f})  \alpha_{k}^\dagger \alpha_{-k}^\dagger
\end{equation}
where $s(\textcolor{black}{h_f})$ and $c(\textcolor{black}{h_f})$ denote trigonometric functions (sine and cosine respectively) of the Bogoliubov angle evaluated at field $\textcolor{black}{h_f}$.   \textcolor{black}{In the remainder of this subsection we suppress all $h_f$ and $t_f$ arguments and indicate explicitly only the initial-time quantities.}  Now, working in the $2\times2$ basis introduced above, we have
\begin{equation}
U \eta^\dagger_k \eta_k U^\dagger =  
\left(
\begin{array}{cc}
U_{11} & U_{12} \\
U_{21} & U_{22}
\end{array}
\right)
\left(
\begin{array}{cc}
s^2 & isc \\
-isc & c^2
\end{array}
\right)
\left(
\begin{array}{cc}
U_{11}^* & U^*_{21} \\
U^*_{12} & U^*_{22}
\end{array}
\right). \label{e:Umateq}
\end{equation}

Using the Bogoliubov transformation at $t_0$ the expression in~\eref{e:Umateq} can then be re-written in terms of the operators $\eta_k(\textcolor{black}{h_0})$ and $\eta_k^\dagger(\textcolor{black}{h_0})$.  It turns out that the only terms coupling $|\tilde{0}(t_0) \rangle$ and $\langle \tilde{0}(t_0)|$  are those proportional to $\eta_ {-k}(\textcolor{black}{h_0}) \eta_ {-k}^\dagger(\textcolor{black}{h_0})$ and the result is
\begin{eqnarray}
\fl \langle \tilde{0}(t_0)| \textcolor{black}{U} \eta_k^\dagger \eta_k \textcolor{black}{U^\dagger}|\tilde{0}(t_0) \rangle& = & c(\textcolor{black}{h_0})^2[s^2|U_{11}|^2+isc U_{11}U_{12}^* - isc U_{11}^* U_{12} + c^2 |U_{12}|^2]  \nonumber \\
&\phantom{=}& +is(\textcolor{black}{h_0})c(\textcolor{black}{h_0})[s^2 U_{21}^*U_{11}+isc U_{11}U_{22}^* - isc U_{12} U_{21}^* + c^2 U_{12}U_{22}^*]  \nonumber \\
&\phantom{=}& -is(\textcolor{black}{h_0})c(\textcolor{black}{h_0})[s^2 U_{11}^*U_{21}+isc U_{12}^*U_{21} - isc U_{11}^* U_{22} + c^2 U_{12}^*U_{22}]  \nonumber \\
&\phantom{=}& +s(\textcolor{black}{h_0})^2[s^2|U_{21}|^2+isc U_{21}U_{22}^* - isc U_{21}^* U_{22} + c^2 |U_{22}|^2] \\
& =& |c(\textcolor{black}{h_0})(s U_{11} - ic U_{12})-is(\textcolor{black}{h_0})(sU_{21}-icU_{22})|^2. 
\end{eqnarray}
If $\mathcal{N}$ defects are generated during the quench then, in the thermodynamic limit where the sum over modes becomes an integral, the defect density is finally given by
\begin{equation}
n_{exc} = \lim_{L\to\infty}\frac{\mathcal{N}}{L} = \frac{1}{\pi} \int_0^\pi  |c(\textcolor{black}{h_0})(s U_{11} - ic U_{12})-is(\textcolor{black}{h_0})(sU_{21}-icU_{22})|^2 \, dk. \label{e:int}
\end{equation}

Crucially, for $\lambda > 1$ the \emph{only} difference is the initial state -- recall that the dynamics is unchanged.  If we start in the current-carrying regime, then  states between $k_-$ and $k_+$ are occupied and since the dynamics of these current-carrying modes is decoupled, we argue that defect production (corresponding to the production of fermion pairs with equal and opposite momenta) can only occur for momenta \emph{outside} this range.  \textcolor{black}{Repeating the calculations leading to~\eref{e:int}, the only difference is a change in the limits of integration so that the result is replaced by}
\begin{eqnarray}
n_{exc}=&\frac{1}{\pi}\int_0^{-k_+}  |c(\textcolor{black}{h_0})(s U_{11} - ic U_{12})-is(\textcolor{black}{h_0})(sU_{21}-icU_{22})|^2 \, dk \nonumber \\
&+ 
\frac{1}{\pi}\int_{-k_-}^\pi  |c(\textcolor{black}{h_0})(s U_{11} - ic U_{12})-is(\textcolor{black}{h_0})(sU_{21}-icU_{22})|^2 \, dk. \label{e:ndefect}
\end{eqnarray}
This integral can be evaluated numerically (see figure~\ref{f:defect}) 
\begin{figure}
\centering
\includegraphics[width=0.8\textwidth]{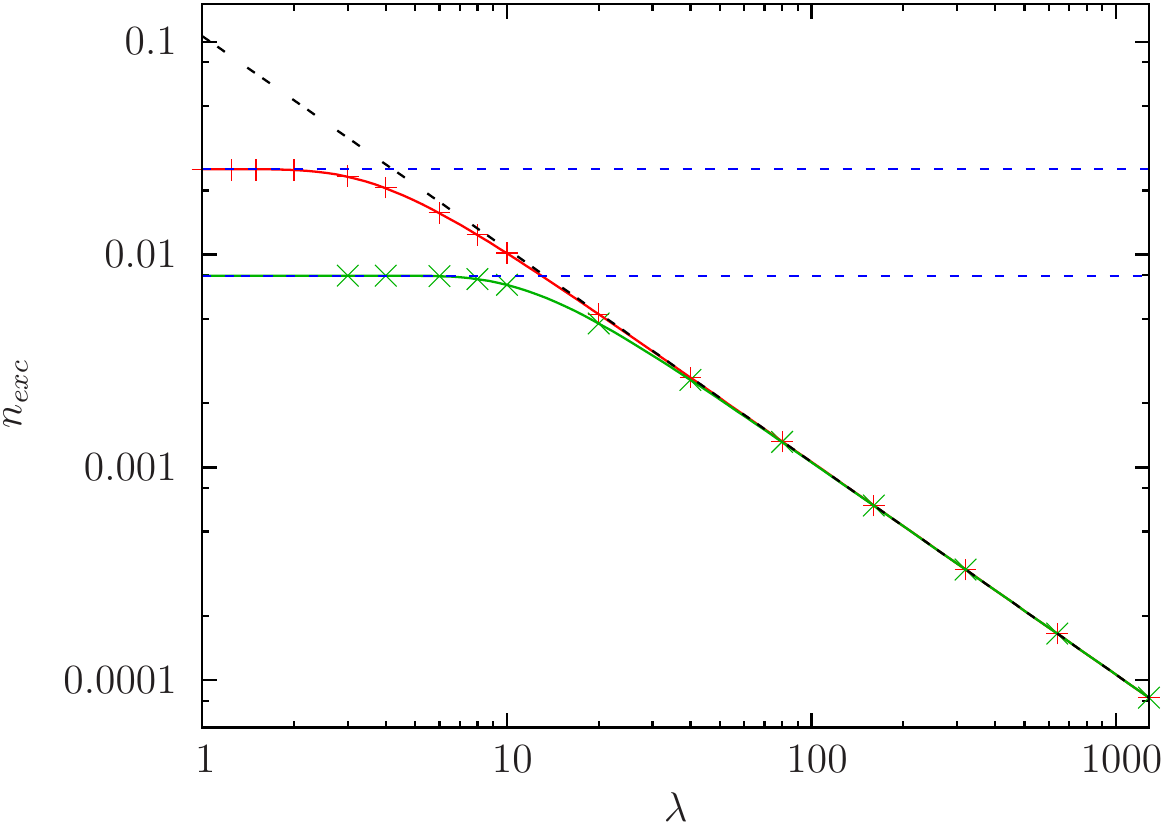}
\caption{Numerical evaluation of~\eref{e:ndefect} for quench protocol $h(t)=2-t/\tau_Q$ from $t_0=-\tau_Q$ to $t_f=\tau_Q$ with $\tau_Q=10$ (red $+$ symbols) and $\tau_Q=100$ (green $\times$ symbols).  The horizontal dashed lines show the small-$\lambda$ scaling prediction of~\eref{e:lamsmall}, the diagonal dashed line is the large-$\lambda$ prediction of~\eref{e:lambig}; the crossover between the regimes is well-described by the error function (solid lines).}
\label{f:defect}
\end{figure}
using the solutions of the differential equations for the matrix elements of $U$ and, significantly, the scaling form can also be predicted by an LZ argument as shown in the next subsection.

\subsection{Mapping to a set of Landau-Zener transitions}
If one restricts the matrices $\tilde{H}_p$ (\ref{e:Hmatrix}) associated to the Ising Hamiltonian $H$ to the non-trivial sector $\{|0\rangle,|2\rangle\}$ 
 the system maps to a set of independent two-level systems each described by the Hamiltonian
\begin{equation}
H(k,t)=[2\cos k + 2h(t)] \; \Id - [\epsilon(t) +b(k)]\; \sigma^z + \Delta(k)\; \sigma^y
\end{equation}
where the $\sigma$'s are Pauli matrices as before, $\Id$ is here the $2\times 2$ identity matrix, and the coefficients are given by $\epsilon(t)=h(t)-2$, $b(k)=4\cos^2(k/2)$, and $\Delta(k)=-2\sin k$.
The instantaneous eigenvalues are 
$[2\cos k + 2h(t)]\pm \sqrt{[\epsilon(t) +b(k)]^2+ \Delta^2(k)}$ associated to the instantaneous eigenvectors $|\pm(t)\rangle$. 
The dephasing factor $b(k)$ can be absorbed by a redefinition of a local time, $t\rightarrow t_k$, for each mode $k$ \cite{Dziarmaga2010}. 

For a driving $\epsilon(t)=-t/\tau_Q$ starting deep in the disordered phase ($h \gg 2$) and ending deep in the ordered phase ($h \simeq 0$),
the main contribution to the excitation density comes from the modes close to the Fermi point $k_F=\pi$ with an excitation probability
\cite{Dziarmaga2010,LZ}
\begin{equation}
p_k=e^{-\pi \tau_Q \Delta^2(k)}= e^{-4\pi \tau_Q  \sin^2 k}\simeq e^{-4\pi \tau_Q  |k-\pi|^2} \; .
\end{equation}
This excitation probability is substantial only for those modes where  $\tau_Q \Delta^2(k)\ll 1$, that is, in a region around the Fermi point of size $|k-k_F|\sim \tau_Q^{-1/2}$ which shrinks towards $k_F$ as the ramping gets slower. 
Since the initial state from which we start the ramping carries a non-vanishing current generated by the population of the negative modes within a region $ [k_-,k_+]$  of the first Brillouin zone, the modes $k\in [k_-,k_+]$ are dynamically protected thanks to the diagonal dynamics (\ref{e:Hmatrix}). No excitation pairs with momenta $\{+k,-k\}$, where $k\in  [k _-,k_+]$, can be generated in the course of time. Consequently, the defect density from the current-carrying initial state is given by
\begin{equation}
n_{exc}=\frac{1}{\pi}\int_0^{-k_+} p_k\;  dk + \frac{1}{\pi} \int_{-k_-}^\pi p_k \; dk \; .
\end{equation}
Since our initial state has a very large value of the transverse field $h$, the boundary mode $-k_+$ is always far away from the Fermi point $k_F=\pi$ and, since the main contribution to $p_k$ comes from the region $|k-k_F|\sim \tau_Q^{-1/2}$, one can simply omit the first integral and write
\begin{equation}
n_{exc}\simeq\frac{1}{\pi} \int_{-k_-}^\pi p_k \; dk \; .
\end{equation}
Plugging  $p_k\simeq e^{-4\pi \tau_Q  |k-\pi|^2}$ in the previous equation, one finally obtains the excitation density
\begin{equation}
n_{exc}\simeq \frac{1}{4\pi \tau_Q^{1/2}} \mathrm{erf}[\sqrt{4\pi \tau_Q} (\pi+k_-)]\; ,
\end{equation}
where $\mathrm{erf}(z)$ is the error function. 
Analysis of the error function presents two limiting cases:
\begin{itemize}

\item If $(8 \pi \tau_Q)^{-1/2} \ll \pi + k_-$, then we have
\begin{equation}
n_{exc} \simeq \frac{1}{4\pi} \tau_Q^{-1/2}, \label{e:lamsmall}
\end{equation}
in agreement with the standard Kibble-Zurek result (remembering here that $\nu=z=d=1$).  Note that, in this limit of slow quenching (large $\tau_Q$, small $\lambda$), there is no dependence on $\lambda$.

\item If $(8 \pi \tau_Q)^{-1/2} \gg \pi + k_-$, then 
\begin{equation}
n_{exc} \simeq \frac{1}{4\pi \tau_Q^{1/2}} \frac{2}{\sqrt{\pi}}\sqrt{4\pi\tau_Q}(\pi+k_-) 
 \simeq \frac{1}{\pi} (\pi+k_-). 
\end{equation}
Furthermore, small $\pi + k_-$ corresponds to $\lambda$ large and expanding (\ref{e:qpm}) in this limit gives
\begin{equation}
\pi+k_- \simeq \frac{1}{\lambda}\left(1-\frac{2}{\textcolor{black}{h_0}}\right)
\end{equation}
where $\textcolor{black}{h_0}$ is the value of the field at the start of the quench.   (Note that, since we start in the ordered state, $\textcolor{black}{h_0}>2$ and the term in the bracket is guaranteed to be positive.)
Hence, we finally get
\begin{equation}
n_{exc} \simeq  \frac{1}{\pi\lambda}\left(1-\frac{2}{\textcolor{black}{h_0}}\right). \label{e:lambig}
\end{equation}
We see that, in this large $\lambda$ limit, there is no dependence on $\tau_Q$ but the initial field $\textcolor{black}{h_0}$ does plays a role since it controls the initial current.  Of course, for a quench starting far away from the critical line we have $\textcolor{black}{h_0} \to  \infty$ and this term drops out.

\end{itemize}

\section{Summary and outlook}
To impose a finite current on the system, we have to populate the vacuum state with modes (current carriers) within a finite set $I_J$ such that the new state is given by $\prod_{k\in I_J} \eta^\dagger_k    |0\rangle$.   Now, if the set $I_J$ has no significant overlap with the critical domain $|k-k_F|\sim \tau_Q^{-1/2} $, the excitation density $n_{exc}$ will be given by the KZM prediction  $n_{exc}\sim   \tau_Q^{-1/2}$ (with $\nu=z=d=1$ for our model). On the contrary, when there is a significant overlap between the two domains, which is exactly what happens at high currents, the excitation density will be lowered.  
At very high current values, the scaling of the defect production is given by $n_{exc}\sim \lambda^{-1}$. The reason is that, in this case, the Lagrange multiplier has to be very high, $\lambda \gg \|H\|$, and the dominant contribution to the effective Hamiltonian $H-\lambda \textcolor{black}{\hat J}$ comes from the current term, the Hamiltonian $H$ itself being a small perturbation.\textcolor{black}{\footnote{\textcolor{black}{The work of~\cite{DoyonBernard} suggests that such large $\lambda$ values can indeed be physically relevant, even for low temperatures.}}} Consequently, the effective spectrum has a dominant contribution of the form $h \lambda \sin k$ and, in the absence of $H$, the ground state is given by occupying all negative modes $k\in [-\pi, 0]$ so there is no possibility of exciting the system: all modes are protected. Now, when we add the Hamiltonian $H$ itself, it will slightly shift the single-particle spectrum by a term $h$ (for $h\gg h_c=2$) resulting in $h \lambda \sin k +h$. Modes close to the Fermi point $k_F=-\pi$ will be unoccupied up to the point $k_-$ where $h \lambda  \sin k_-+h=0$, that is, to leading order up to $k_-=-\pi+1/\lambda$. We are then left close to $k_F=-\pi$ with an unoccupied domain of size $|k-k_F|=1/\lambda$ that can be excited during the quench, leading finally to $n_{exc}=\frac{1}{\pi} \int_{\pi -1/\lambda}^\pi p_k dk= \frac{1}{\pi} \int_{\pi -1/\lambda}^\pi dk= \frac{1}{\pi\lambda}$. At lower fields, the same argument (linearizing the dispersion relation close to $k_F=-\pi$) will lead to $k_-\simeq -\pi+ \frac{1}{\lambda}\left(1-\frac{2}{\textcolor{black}{h_0}}\right)$ and consequently to 
 $n_{exc}\simeq   \frac{1}{\pi\lambda}\left(1-\frac{2}{\textcolor{black}{h_0}}\right)$. 

In summary, we have studied the defect generation when driving an Ising quantum chain through its critical point from an initial state that carries a net energy current. The current value is imposed via a Lagrange multiplier field. 
We have shown that low values of the energy current do not affect the density of defects that are generated during the crossing of the critical point. In contrast, at large enough currents the current carriers are dynamically protected leading to a significant suppression of the defect generation.

It would be very interesting to test the influence of an initial current of particles on the generation of topological defects in models which do not reduce to a set of free particles and where, consequently, the generation of defects does not reduce to a Landau-Zener problem.  A potential candidate to probe that influence is the non-integrable Bose-Hubbard model with a slow driving from the Mott insulator phase to the superfluid phase. In the zero-current situation there are already some analytical KZM predictions based on the fact that the $d$-dimensional Mott insulator to superfluid transition belongs to the universality class of the $(d+1)$-dimensional $XY$ spin model \cite{Dziarmaga2010}. The possibility of extending these KZM predictions to the case of a current-carrying initial state, with $(d+1)$-dimensional classical analogue, is currently under consideration.

\ack
RJH is grateful for the hospitality of the Groupe de Physique Statistique in the Institut Jean Lamour, Nancy where this work was carried out.  
The authors would like to thank Mario Collura for useful discussions.

\end{document}